\documentclass[11pt,a4paper,aps,preprintnumbers]{article}
\usepackage{jcappub}

\usepackage{dcolumn}
\usepackage{epsfig}
\usepackage{bbm}
\usepackage{bbding}
\usepackage{amsmath}
\usepackage{amsfonts}
\usepackage{amssymb}
\usepackage{mathtools}
\usepackage{dsfont}
\usepackage{bm}
\usepackage{float}
\usepackage{graphicx}
\usepackage{subcaption}
\usepackage{psfrag}
\usepackage{hyperref}
\usepackage[capitalise]{cleveref}
\usepackage{tikz}
\usetikzlibrary{positioning,arrows.meta}
\usepackage{pgfplots}
\usepackage{pgfplotstable}
\usepackage{subfiles}
\usepackage{tabularx}
\usepackage{ltablex}
\usepackage{booktabs}
\usepackage[export]{adjustbox}
\usepackage{listings}
\usepackage{multirow} 
\usepackage{cancel,slashed}
\usepackage{bbm}
\usepackage{graphicx}
\usepackage{latexsym}
\usepackage{transparent}
\usepackage{mathtools}
\usepackage{array}
\usepackage{makecell}
\usepackage{graphics,psfrag}
\usepackage{placeins}
\usepackage{nowidow}
\usepackage{listings}
\usepackage[normalem]{ulem}
\usepackage{environ}
\usepackage{enumitem}  
\usepackage{soul}

\def\la{~\mbox{\raisebox{-.6ex}{$\stackrel{<}{\sim}$}}~}
\def\ga{~\mbox{\raisebox{-.6ex}{$\stackrel{>}{\sim}$}}~}
\newcommand{\mpl}{M_{\rm Pl}}

\graphicspath{{Figures/}}

\begin{document}

\title{CMB spectral distortions from enhanced primordial perturbations: the role of spectator axions}

\author[a,b]{Margherita Putti,}
\author[c,d,e]{Nicola Bartolo,} 
\author[c,d]{Sukannya Bhattacharya,}
\author[c,d]{Marco Peloso}

\affiliation[a]
{Deutsches Elektronen-Synchrotron DESY, Notkestr. 85, 22607 Hamburg, Germany}
\affiliation[b]{II. Institut f\"{u}r Theoretische Physik, Universit\"{a}t Hamburg
Luruper Chaussee 149, 22607 Hamburg, Germany}
\affiliation[c]{Dipartimento di Fisica e Astronomia “Galileo Galilei”, Universit\`a di Padova, 35131 Padova, Italy}
\affiliation[d]{INFN, Sezione di Padova, 35131 Padova, Italy}
\affiliation[e]{INAF - Osservatorio Astronomico di Padova, vicolo dell’ Osservatorio 5, I-35122 Padova, Italy}

\emailAdd{margherita.putti@desy.de}
\emailAdd{nicola.bartolo@pd.infn.it}
\emailAdd{sukannya.bhattacharya@unipd.it}
\emailAdd{marco.peloso@pd.infn.it}

\abstract{Primordial scalar and tensor modes can induce Cosmic Microwave Background  spectral distortions during horizon re-entry. We investigate a specific mechanism proposed for this purpose, characterized by the coupling of an SU(2) gauge field to an axion undergoing a momentary stage of rapid evolution during inflation. Examining in details the perturbations produced by this model, we find that spectral distortions from the scalar modes significantly dominate those arising from the tensors. This holds true also for an earlier version of the model based on a U(1) gauge field. The scalar-induced distortions might be observed in future experiments, and the current COBE/FIRAS constraints already limit the parameter space of these models. Additionally, we find that delaying the onset of fast roll in the SU(2) scenario (to enhance the modes at the scales relevant for spectral distortions, while respecting the CMB constraints at larger scales) poses a greater challenge compared to the U(1) case. We propose a way to control the axion speed by varying the size of its coupling to the gauge fields. 
}

\maketitle


\section{Introduction}

Since the measurements taken by COBE/FIRAS~\cite{Mather:1993ij} in the $1990$s, the Cosmic Microwave Background (CMB) radiation is a remarkable example of a nearly perfect blackbody spectrum at the temperature $T_0=(2.726\pm0.001)K$ today~\cite{Fixsen:2009ug}, representing a cornerstone in modern cosmology. This striking agreement arises from the thermal equilibrium between matter and radiation during the early stages of the universe. However, as the universe evolves, various mechanisms come into play, potentially leading to modifications in the CMB frequency spectrum. The thermalization process of the CMB in the early universe has been extensively studied. In the early stages (redshift $z\geq 2 \times 10^{6}$), a blackbody spectrum is maintained through ongoing processes, such as Compton scattering, Bremsstrahlung, and double Compton scattering. As the universe expands, these interactions become less efficient, allowing for deviations from the blackbody spectrum. These deviations, known as CMB spectral distortions (SDs), are induced by energy injections occurring at epochs with $z\leq 10^6$, and therefore  represent a yet unexplored new window into both
the early and the late universe physics. A guaranteed mechanism of early SDs, predicted by the standard $\Lambda$CDM cosmological model, is that due to the so called Silk-damping effect \cite{Silk:1967kq}. Additionally, there are other numerous processes, occurring at a variety of redshifts, known to potentially disturb the thermal distribution, including reionization and structure formation \cite{Chluba:2012gq}, energy injection from annihilating or decaying particles \cite{Hu:1992dc}, the presence of small-scale primordial magnetic fields \cite{Jedamzik:2018itu}, and cosmic strings \cite{Ostriker&Thompson}, among others. These diverse mechanisms provide intriguing avenues to investigate and understand the origins of SDs as signatures of new physics. Also, spatial anisotropies 
of CMB SDs (and their cross-correlations with CMB temperature and polarization fields) provide further ways to exploit SDs as a valuable observable to have a new insight into both early and late time physics (including, but not limited to primordial non-Gaussianity), see, e.g.,~\cite{CORE:2017krr,Pajer:2012vz,Ganc:2012ae,Emami:2015xqa,Chluba:2016aln,Ravenni:2017lgw,Bartolo:2015fqz,Shiraishi:2016hjd,Remazeilles:2021adt,Rotti:2022lvy,Chluba:2022efq,Kite:2022eye, Pajer:2012qep}.

CMB distortions are classified into two main types based on the epoch of the energy release: $\mu$-type and $y$-type distortions. The $\mu$-distortion exhibits a frequency dependent chemical potential, and it is generated after the decoupling of the double Compton scattering ($z\sim 10^6$), while the Compton scattering is still active, guaranteeing kinetic equilibrium. Conversely, the $y$ distortion is produced as the Compton scattering is no longer efficient in maintaining this equilibrium ($z\sim 10^5$). Measurement of these CMB distortions provides a powerful tool for investigating the thermal history of the Universe.
The most precise measurement of the CMB spectrum to date is provided by COBE/FIRAS, which constrained the distortions to $\mu\leq 9\times 10^{-5}$ and $y\leq1.5\times10^{-5}$ at a $95\%$ C.L.~\cite{Fixsen:1996nj}. A later analysis by \cite{Bianchini:2022dqh} improved the $\mu$-distortion constraint to $\mu\leq 4.7 \times 10^{-5}$ at a $95\%$ C.L., thanks to an improved component separation and Bayesian analysis approach. The last decade has witnessed an intense ongoing discussion regarding potential future missions (PIXIE~\cite{Kogut:2010xfw}, PRISM~\cite{PRISM:2013fvg}, COSMO~\cite{Masi:2021azs}, BISOU~\cite{Maffei:2021xur}) that could detect $\mu-$ and $y-$distortions down to the $10^{-9} - 10^{-8}$ range~\cite{Chluba:2019kpb,Chluba:2019nxa}. 

Interestingly, CMB spectral distortion observations would let us venture into much smaller scales than those available by CMB anisotropy measurements, which range from our horizon scale $k\simeq 2\times 10^{-4}$~Mpc$^{-1}$ to $k\simeq0.2$~Mpc$^{-1}$.
Indeed, any scalar fluctuation in the CMB temperature will be erased by Silk damping around the dissipation scale $k_D$, which, for $z\simeq 2\times 10^6$, is of the order $k_D\simeq 2\times 10^4$~Mpc$^{-1} \gg 0.2$~Mpc$^{-1}$~\cite{Silk:1967kq}. Even in conventional cosmology (for example, even without assuming entropy injection from some long-lived species), entropy release associated to cosmological perturbations produced during inflation and that re-enter the horizon at $z < {\rm O } \left( 10^6 \right)$, generates distortion. Under assumption of (approximate) scale invariance of primordial perturbations, a curvature power spectrum of $P_\zeta\sim 2\times 10^{-9}$ guarantees a signal of $\mu,y\sim 10^{-8}$, that could be a target for Voyage 2050~\cite{Chluba:2019nxa}. Moreover, improving on the experimental results can
play a vital role in constraining various inflationary models that involve a breaking of scale invariance after CMB scales, leading to enhanced power at smaller scales than the CMB. 
 
Inflation, however, gives rise not only to scalar modes, but also to tensor modes (Gravitational Waves, GWs). Recent detections of GWs by the ground-based LIGO~\cite{LIGOScientific:2014pky}, Virgo~\cite{VIRGO:2014yos}, and KAGRA \cite{KAGRA:2020tym} (LVK) observatories, as well as evidence from the pulsar timing arrays (PTA) \cite{Antoniadis:2023rey,NANOGrav:2023gor}, have expanded the field of GWs, leading to significant advancements in theory and observation. With the advent of new experiments, from PTA to astrometry~\cite{Moore:2017ity}, from laser interferometer space antennas (LISA)\cite{LISA:2017pwj} to new ground based detectors (such as Einstein Telescope~\cite{Punturo:2010zz} and Cosmic explorer \cite{Evans:2021gyd}) we are entering an era of unprecedented opportunities to explore the cosmos through gravitational wave signals. 
It has been shown that CMB SDs can also play a part in the observation of GWs \cite{Chluba:2014qia,Kite:2020uix}. Indeed, tensor modes can generate SDs at their horizon re-entry, just like scalar modes.  The effect of this contribution to SDs has been quantified in~\cite{Chluba:2014qia}.
The authors of \cite{Ramberg:2022irf} investigated the spectral distortions from tensor and scalar fluctuations arising in post-inflationary scenarios that lead to production of GWs, finding that the contribution from the tensor modes were highly suppressed with respect to the scalar ones.  

In this paper we study the SDs from scalar and tensor modes in a specific context, namely in inflation models where rolling axions are coupled to gauge fields. This includes one of the scenarios considered in previous works by \cite{Kite:2020uix}, that focused on the tensor-induced SDs. As we show below, our results agree with~\cite{Kite:2020uix}. In addition to the generation of tensor modes leading to CMB distortions in these models, we also investigate the generation of scalar modes leading to CMB distortions. The latter effect is generally present, as scalar modes are generated gravitationally, along with tensor modes. In fact, most models produce a significantly higher scalar signal as compared to the tensor one, making it qualitatively inevitable and in need of quantitative assessment. The production of scalar modes gives a higher hope for a detection of SDs or a possibility to constrain these model, compared to relying only on tensor-induced SDs. 
The issue of scalar production that limits the detectable signatures of tensor production has been a long-standing concern in various mechanisms that generate additional tensor modes beyond the standard quantum-vacuum tensor fluctuations. For instance, models designed to enhance the tensor-to-scalar ratio $r=\frac{P_T}{P_\zeta}$ at CMB scales by generating an additional contribution of gravitational waves also unavoidably produce scalar modes. This not only affects the denominator of the ratio, but also can encounter strong bounds from the non-Gaussian character of the sourced scalar modes~\cite{Barnaby:2012xt,Biagetti:2014asa}. Moreover, the production of observable GWs, as might be detected by future missions like LISA or the LVK collaboration, encounters a delicate trade-off: if too many fields are involved in generating the GWs, the excessive energy injection can lead to the formation of primordial black holes (PBHs) at undesired levels \cite{Kalaja:2019uju,Kohri:2014lza}. As such, careful considerations of both scalar and tensor modes are essential, as their coexistence necessitates a comprehensive evaluation of their mutual effects on cosmological observables.

Firstly, we explore a $U(1)$ model designed~\cite{Namba:2015gja} to maximize the relative production of tensor versus scalar modes. This is done by confining the rolling axion and gauge fields to a ``spectator sector'', decoupled as much as possible from the inflaton sector from which the observable scalar perturbations originate. By avoiding any direct coupling in the potential, the two sectors are coupled only by the unavoidable gravitational interactions. Although this can result in an increased tensor signal, while suppressing to a minimum level the production of scalar modes, we find that the tensor contribution to SDs is still significantly smaller than the scalar one. Therefore, isolating and probing the much subdominant tensor contribution to SDs is extremely challenging. 

Next, we turn our attention to the $SU(2)$ version of this model, introduced in~\cite{Dimastrogiovanni:2016fuu}, and already discussed in~\cite{Kite:2020uix} as a concrete model for SDs from tensor modes. The linearized study of the primordial perturbations leads to the conclusion that the production of additional tensor modes in this model can be significantly greater than that of additional scalars. This is due to the existence of an unstable gauge field mode, that experiences significant growth while the axion is rolling, and that couples linearly to GWs but not to the curvature perturbation $\zeta$. However, the unstable gauge mode sources a significant amount of scalar perturbations nonlinearly~\cite{Peloso:2016gqs}. We find that also in this model the SDs generated by tensor modes are much smaller than those due to the nonlinearly produced scalar perturbations. 
We also find that generating a localized signal in the primordial tensor and scalar perturbations in the SU(2) model is more challenging than in the U(1) case.~\footnote{This can only be seen from evolving the background equations for the model (with parameters appropriately chosen so to generate a roll of the axion only for a limited amount of e-folds during inflation), and not from the parametrization of the peaked signal adopted in some literature~\cite{Thorne:2017jft}.} This is due to additional interactions present in the SU(2) case, that make it more difficult to keep the gauge field/axion sector in a `dormant' phase, so to obtain a fast axion roll at some specific desired time. We present a possible way out to this problem, by relating the onset of the fast axion roll to a sudden change of the coupling between the axion and the gauge fields. 

The plan of the work is the following. In \cref{SD}, we present a short overview of the SDs arising from primordial perturbations, summarizing results present in the literature, and providing explicit expressions for both the scalar and tensor window functions (that encode the SDs arising from these fluctuation modes). In \cref{U1} we review the $U(1)$ spectator model, and we calculate the distortions resulting from the enhanced power spectra predicted by this model. In \cref{SU} we shift our focus to the $SU(2)$ spectator model. We revisit the dynamics of the model, in particular concerning the generation of a bump in the power spectra. Once we are able to generate this enhancement, we then compute the SDs. 
Our findings are then reviewed in the concluding Section~\ref{sec:conclusions}.

\section{Spectral distortions from primordial perturbations}
\label{SD}

Photons in the early universe have the Bose-Einstein distribution function 
\begin{equation}
f \left( x \right) = B \left( x \right) \equiv \frac{1}{{\rm e}^x-1} \;\;,\;\; x \equiv \frac{p}{T} \;, 
\label{BB}
\end{equation} 
where $p$ is the physical momentum of the photons and $T$ their temperature. The expansion of the universe modifies this distribution function through a rescaling of the temperature $T$, so that the primordial photons retain this blackbody spectrum in absence of interactions. Such interactions, and, in general, energy transfers to or from the photon field will cause a distortion of this spectrum~\cite{Zeldovich:1969ff,Sunyaev:1970er,Burigana:1991eub,Hu:1992dc}. To linear order in the parameters $\Delta T/T ,\, \mu ,\, y$, the modifications are typically of the form~\footnote{We refer the readers to ref.~\cite{Lucca:2019rxf} for a pedagogical derivation of the expressions reported here. For a more detailed study of modifications beyond the modelization in eq.~(\ref{distortions}) see for instance refs.~\cite{Chluba:2011hw,Khatri:2012tw,Chluba:2013vsa,Chluba:2015hma}.}
\begin{equation}
\Delta f \left( x \right) = \frac{\Delta T}{T} \, G \left( x \right) + \mu \, M \left( x \right) + y \, Y \left( x \right) \;.  
\label{distortions}
\end{equation}
Only the last two terms in this expression represent a  distortion of a blackbody distribution. The first term, 
\begin{equation} 
G \left( x \right) \equiv - x \, \frac{\partial B \left( x \right)}{\partial x} = \frac{x \, {\rm e}^x}{\left( {\rm e}^x - 1 \right)^2} \;, 
\end{equation}
represents instead a change of temperature of the photon distribution without altering its blackbody shape. This is the only modification introduced by any entropy injection that occured at early redshifts, $z \ga z_{\rm dc} \equiv 2 \times 10^6$~\cite{Burigana:1991eub,Hu:1992dc}, when processes that change the number (double Compton and bremsstrahlung) and energy (Compton scattering) of photons requilibrate the distortion into a new blackbody spectrum with a modified temperature. Below this redshift, and up to approximately the redshift $z \equiv z_{\mu y} = 6 \times 10^4$~\cite{Hu:1992dc,Chluba:2015bqa}, double Compton and bremsstrahlung processes progressively go out of equilibrium, while Compton scattering is still active and guarantees kinetic equilibrium. Any entropy release then results in both a change of temperature and of chemical potential of the photon distribution, with the latter effect given by 
\begin{equation}
M \left( x \right) \equiv - G \left( x \right) \left( \frac{1}{x} - 0.4561 \right) \;. 
\end{equation} 
For an entropy injection below this redshift, and up to recombination ($z = z_{\rm rec} \simeq 1100$), Compton scatterings of the photons with the electrons (assumed to have a Maxwellian phase-space distribution) take place with a progressively decreased efficiency, giving rise to a so called Compton y-distortion, with shape 
\begin{equation}
Y \left( x \right) \equiv G \left( x \right) \left( x \, \frac{{\rm e}^x+1}{{\rm e}^x -1} -4 \right) \;. 
\end{equation} 

To linear order, an entropy release $\dot{Q}$ (where $Q \left( z \right)$ is the energy density injected at the redshift $z$ in the thermal bath, and dot denotes derivative with respect to time) produces  distortions that can be parametrized as 
\begin{equation}
\mu = 1.4  \int_0^\infty {\tilde {\cal J}}_\mu \left( z \right) \frac{d \left( Q / \rho_\gamma \right)}{d z} \, d z \;\;,\;\; 
y = \frac{1}{4} \int_0^\infty {\tilde {\cal J}}_y \left( z \right) \frac{d \left( Q / \rho_\gamma \right)}{d z} \, d z  \;, 
\label{mu-y}
\end{equation}
where $\rho_\gamma$ is the energy density of the thermal distribution (without the injection) and the two visibility functions ${\tilde {\cal J}}_{\mu,y}$ account for the particle physics processes that we have summarized above. 

The visibility functions have been studied with increasing accuracy in the literature, see for instance the discussion in ref.~\cite{Chluba:2016bvg}. The simplest approach is to take two square top hat functions evaluating to $1$ in the redshift intervals $z_{\mu y} < z < z_{\rm dc}$ (for the $\mu-$distortion) and $z_{\rm rec} < z < z_{\mu y}$ (for the $y-$distortion) and to $0$ outside these intervals. Improvements have been obtained by noting that the efficiency does not abruptly vanish at $z_{\rm dc}$ ~\cite{Sunyaev:1970er,Danese,Burigana:1991eub,Hu:1992dc}, which can be accounted for by a ${\rm e}^{-\left( z / z_{\rm dc} \right)^{5/2}}$ factor, and by considering the fact that the
transition between $\mu-$ and $y-$distortions is not abrupt at $z_{\mu y}$. We employ the analytic approximation \cite{Chluba:2015bqa}
\begin{align}
y \simeq \frac{1}{4}\int_{z_{\rm rec}}^{\infty} dz'  \text{e}^{-(z'/z_{dc})^{5/2}} \frac{d(Q/\rho_\gamma)}{dz'}\mathcal{J}_y(z')
\;\;\;,\;\;\; 
\mu \simeq 1.4 \int_{z_{\rm rec}}^{\infty}dz'  \text{e}^{-(z'/z_{dc})^{5/2}}  \frac{d(Q/\rho_\gamma)}{dz'}\mathcal{J}_\mu(z')\,, 
\label{dist}
\end{align}
with 
\begin{equation}
\mathcal{J}_{y}(z) \simeq 
\left(1+\left[\frac{1+z}{z_{\mu y}}\right]^{2.58}\right)^{-1} \;\;,\;\; 
\mathcal{J}_\mu(z) \simeq 1-\mathcal{J}_y(z) \, . 
\end{equation}

Let us discuss the entropy injections of interest for our work. A guaranteed source of distortions is from the energy release due to  the Silk-damping~\cite{Silk:1967kq} of primordial small-scale
perturbations after their horizon re-entry~\cite{Sunyaev:1970plh,Daly:1991uob,barrow1991primordial,Hu:1994bz}. This gives rise to CMB SDs that are directly related to the shape and amplitude of the primordial scalar power spectrum $P_\zeta$. Defining the latter as in eq. (\ref{P-zeta}),~\footnote{This definition corresponds to the one of ref.~\cite{Kosowsky:1995aa} times a $\frac{k^3}{2 \pi^2}$ factor. In our notation, a scale-invariant power spectrum corresponds to a constant $P_\zeta$. The same applies to the tensor power spectrum that we consider below. We only consider the effect of an adiabatic primordial perturbation. Ref.~\cite{Chluba:2013dna} studied the distortions due to scalar isocurvature modes.} the injection of energy from the scalar modes can be approximated as~\cite{Chluba:2015bqa} (see also refs.~\cite{Chluba:2012gq,Chluba:2012we,Lucca:2019rxf}) 
\begin{equation}
\frac{d(Q/\rho_\gamma)}{dz}\bigg|_\zeta\simeq - 3.25 \int d \ln k \, P_\zeta \left( k \right) \sin^2 \left( k r_s \right) \, \partial_z \, {\rm e}^{-2 k^2/k_D^2} \;, 
\label{Q-zeta}
\end{equation} 
where, assuming radiation domination, the sound horizon is $r_s \simeq c_s \, \eta \simeq 2.7 \times 10^5 \left( 1 + z \right)^{-1}$, while the damping scale is $k_D \simeq 4.0 \times 10^{-6} \left( 1 + z \right)^{3/2} {\rm Mpc}^{-1}$.~\footnote{In writing the numerical coefficient of eq.~(\ref{Q-zeta}) and the following relations for the heating rates and window functions taken from the literature we set the
fractional contribution of neutrinos to the energy density
of relativistic species to $R_\nu \equiv \rho_\nu / \left( \rho_\gamma + \rho_\nu \right) \simeq 0.41$.} 

\begin{figure} 
\centering
\includegraphics[scale=0.6]{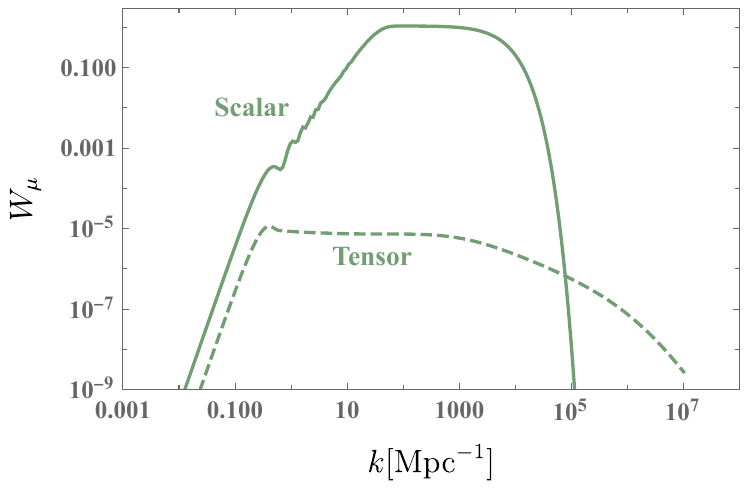}
\includegraphics[scale=0.6]{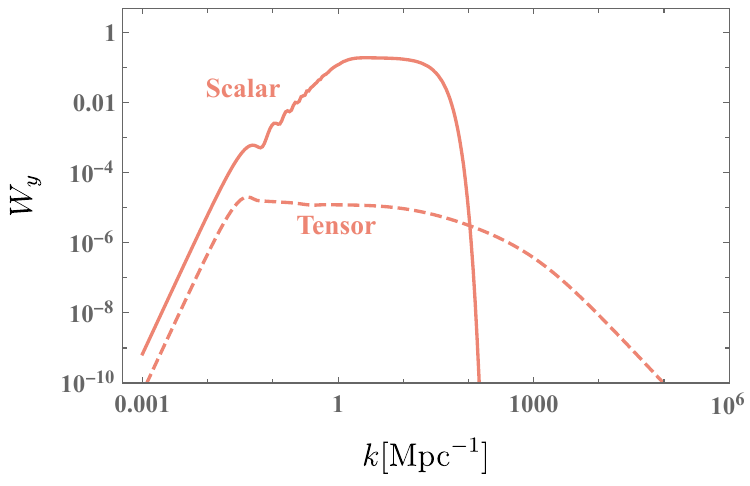}
\caption{Window functions for (left) $\mu-$ and (right) $y-$distortions. They encode the effect of primordial modes on the distortions as in eq.~(\ref{mu-y-zeta}). The solid (respectively, dashed) line corresponds to the scalar (respectively, tensor) window function.}
\label{fig:Wcomparison_ap}
\end{figure}

Combining eqs. (\ref{mu-y}) and (\ref{Q-zeta}), it is customary to write the distortions as 
\begin{equation} 
\mu_\zeta = \int_{k_{\rm min}}^\infty d \ln k \, P_\zeta \left( k \right) \,  W_\mu^\zeta \left( k \right) \;\;,\;\; 
y_\zeta = \int_{k_{\rm min}}^\infty d \ln k \, P_\zeta \left( k \right) \,  W^y_\zeta \left( k \right) \;, 
\label{mu-y-zeta}
\end{equation}
with $k_{\rm min} \simeq 1 \, {\rm Mpc}^{-1}$. We show the scalar window functions obtained in this way with solid lines in Figure \ref{fig:Wcomparison_ap}.~\footnote{Analytic approximations for the scalar window functions can be found in Ref.~\cite{Chluba:2013dna}.}
%
%

Primordial tensor modes can also source CMB distortions~\cite{Ota:2014hha}. Contrary to scalar modes, they create a local quadrupole anisotropy without the need of photon diffusion. Scatterings between photons and electrons in presence of this anisotropy then cause nearly scale independent dissipation~\cite{Ota:2014hha,Chluba:2014qia},  giving rise to the distortions. Ref.~\cite{Chluba:2014qia} provided analytical relations for the heating rate from tensors and for the resulting distortions. The heating rate can be expressed as  
\begin{equation}
   \frac{d (Q/\rho_\gamma)}{dt}\bigg|_T\simeq \frac{4H^2}{45 \dot{\tau}}\int_0^\infty d \ln k \, P_T \left( k \right) \, \mathcal{T}_\theta \left( \frac{k}{a \dot{\tau}} \right) \mathcal{T}_h \left( k \eta \right) \text{e}^{-\Gamma_\gamma\eta} \,, 
   \label{Q-T}
\end{equation}
where $\dot{\tau}=\sigma_T N_e=4.4 \times 10^{-21}(1+z)^3 \mathrm{~s}^{-1}$ is the Thomson scattering rate, $\eta$ is conformal time, and $a$ is the scale factor. In this relation, $P_T \left( k \right)$ is the tensor power spectrum (summed over the two polarizations), defined analogously to eq. (\ref{P-zeta}). The gravitational wave transfer function~\cite{Dicus:2005rh}
\begin{equation}
\mathcal{T}_h \left( x \right) \simeq 2\left\{\sum_{n=0,even}^6 a_n[n j_n(x) -x j_{n+1}(x)]\right\}^2\,,
\end{equation}
(where the $j_n$ are the spherical Bessel functions, and $a_0=1$, $a_2= 0.243807$, $a_4=5.28424\times 10^{-2}$, $a_6=6.13545\times 10^{-3}$) accounts for the damping due to neutrino free streaming~\cite{Weinberg:2003ur,Watanabe:2006qe}, while the rate $\Gamma_\gamma$ accounts for the damping due to photons, and $\Gamma_\gamma \, \eta \simeq 5.9 \, a$ as derived in eq. (D.6) of~\cite{Chluba:2014qia}. Finally, the photon transfer function encodes how the distortion is produced by the tensor mode, and it can be approximated by~\cite{Chluba:2014qia}
\begin{equation}
\mathcal{T}_\theta \left( x \right) \simeq \frac{1+4.48 x+91 x^2}{1+4.64x+90.2x^2+100 x^3+55.0x^4}\,.
\end{equation}

From these expressions, we can express the distortions analogously to eq.~(\ref{mu-y-zeta}), namely 
\begin{equation} 
\mu_T = \int_0^\infty d \ln k \, P_T \left( k \right) \,  W_\mu^T \left( k \right) \;\;,\;\; 
y_T = \int_0^\infty d \ln k \, P_T \left( k \right) \,  W^y_T \left( k \right) \;, 
\label{mu-y-T}
\end{equation}
in terms of the tensor power spectrum and the tensor window functions 
\begin{align}
W_y^T&\simeq\int dz \frac{1}{(1+z)} \frac{H}{45 \dot{\tau}}  \mathcal{T}_h(k\eta) \, \mathcal{T}_\theta\left(\frac{k}{\tau'}\right) \text{e}^{-\Gamma_\gamma\eta}\,\mathcal{J}_t(z) \,, \nonumber\\
W_\mu^T&\simeq \int dz\,\text{e}^{-(\frac{z}{z_{dc}})^{\frac{5}{2}}} \frac{1}{(1+z)} \frac{5.6H}{45 \dot{\tau}} \mathcal{T}_h(k\eta)\, \mathcal{T}_\theta\left(\frac{k}{\tau'}\right) \text{e}^{-\Gamma_\gamma\eta}\, \mathcal{J}_\mu(z)
\,.
\label{F_mu_y_T}
\end{align}
that we show with dashed lines in \cref{fig:Wcomparison_ap}.

The comparison between the tensor and scalar window functions shown in this figure allows for the following  observation: while the scalar window functions are significantly greater than the tensor ones at large scales, the tensor window functions become greater than the scalar ones at smaller scales, approximately at $k \gtrsim 7.5 \times 10^4$ Mpc$^{-1}$ for $\mu-$distortions and at $k \gtrsim 10^2$ Mpc$^{-1}$ for $y$-distortions. This phenomenon aligns with expectations, considering that the scalar window function starts to decrease around the diffusion scale due to Silk damping effects. In contrast, as already mentioned, tensor modes directly induce a quadrupole anisotropy, and a consequent CMB distortion, without relying on photon diffusion. Accordingly, the damping process for tensor modes is significantly less scale-dependent, as the figure shows. When comparing the effects of scalar vs. tensor modes it is also important to recall that the window functions multiply the primordial scalar and tensor power spectra, and that tensor modes are typically produced with a significantly lower amplitude than scalars in vanilla slow-roll inflation. Therefore, in hope to have a visible distortion from tensors (which is not overshadowed by that from scalars), one needs to resort to specific inflationary models that can generate enhanced tensor perturbations at the scales relevant for the distortions~\cite{Kite:2020uix}. In the following, we consider one class of models that have been proposed in the literature for this purpose, namely enhanced tensor perturbations from gauge fields amplified by a rolling axion during inflation. In addition to the existing literature, we consider also the effect of scalars that are also unavoidably produced by the gauge fields. 

\section{Rolling spectator axion coupled to a U(1) gauge field}
\label{U1}
As already mentioned in the Introduction, one of the models for the study of SDs from tensor perturbations considered in the literature~\cite{Kite:2020uix} is characterized by a spectator axion field $\chi$ that rolls for a few e-folds during inflation, sourcing an SU(2) field at the scales that leave the horizon in those e-folds~\cite{Thorne:2017jft}. These gauge fields source a localized peak of gravitational waves, which might in turn give rise to significant CMB SDs~\cite{Kite:2020uix}. The mechanism of~\cite{Thorne:2017jft} is a variation of that at the basis of chromo-natural inflation~\cite{Adshead:2012kp} , in which the rolling axion is identified with the inflaton, and the production of gauge fields continuously takes place during inflation.\footnote{A spectator axion coupled to an SU(2) gauge field was also considered in ref.~\cite{Dimastrogiovanni:2016fuu}, in which the axion rolls continuously throughout inflation.} We will review this application in the next section. 

Here, we study the application for SDs of the analogous (and earlier) model of~\cite{Namba:2015gja},~\footnote{SDs in this model were considered in association to primordial scalar perturbations that can result in primordial black hole dark matter~\cite{Garcia-Bellido:2017aan}.} in which the rolling spectator axion is coupled to a U(1) gauge field. Beside the rolling axion $\chi$ and the $U(1)$ gauge field $A_\mu$, the model also contains an inflaton $\phi$, and it is characterized by the Lagrangian 
\begin{equation}
    \mathcal{L}=-\frac{1}{2}(\partial \phi)^2-\frac{1}{2}(\partial \chi)^2- V(\phi)+U(\chi)-\frac{1}{4}F_{\mu \nu}F^{\mu \nu}-\frac{\lambda}{4 f} \chi F_{\mu \nu}\tilde{F}^{\mu\nu}\,, 
    \label{model-U1}
\end{equation}
where $F_{\mu \nu} \equiv \partial_\mu A_\nu - \partial_\nu A_\mu$ is the U(1) field strength, while  $\tilde{F}^{\mu \nu} \equiv \frac{1}{2} \frac{\epsilon^{\mu \nu \alpha \beta}}{\sqrt{-g}}F_{\alpha \beta}$ is its dual, with $\epsilon^{\mu \nu \alpha \beta}$ totally anti-symmetric (and $\epsilon^{0123}=+1$) and $g$ the determinant of the metric.~\footnote{We assume a FLRW geometry, with line element $ds^2 = - dt^2 + a^2 \left( t \right) d \vec{x}^2$, where $a$ is the scale factor and $H \equiv \frac{\dot{a}}{a}$ the Hubble rate (dot denotes a derivative with respect to physical time).} The coupling of the axion to the gauge field has been parametrized in terms of the (mass dimension one) axion scale $f$ and the (dimensionless) coefficient $\lambda$. The axion potential is chosen to have the typical cosine dependence~\footnote{We instead do not specify the inflaton potential $V \left( \phi \right)$ in this section, and we treat $H$ as a constant during inflation, since a typical slow-roll variation of $H$ impacts in a negligible way the evolution of the axion in the U(1) case. A specific inflaton potential is considered in the SU(2) case studied in the next section, since in that context the background evolution is less trivial.} 
\begin{equation}
    U(\chi)= \Lambda^4 \, \left[\cos\left(\frac{\chi}{f}\right)+1\right]\,.  
\label{Vaxion-cos}
\end{equation}
If the curvature $m^2$ of this potential is tuned to be of the order of the square of the Hubble rate $H^2$ during inflation, and if the axion is initially close to a local maximum of the potential (say, for definiteness, $\chi \simeq 0^+$ initially), then the axion rolls to a minimum of the potential (in this case, to $\chi = f \pi$) in a few efolds $\delta N \simeq \frac{1}{\delta}$, where~\cite{Namba:2015gja}  
\begin{equation}
\delta \equiv \frac{m^2}{3 \, H^2} \equiv \frac{\Lambda^4}{3 H^2 f^2} \;. 
\label{delta}
\end{equation} 
During its brief roll, the axion sources gauge fields, with an amplitude that is exponentially proportional to the parameter $\xi \equiv \frac{\lambda \dot{\phi}}{2 f H}$. In the following, we denote with $\xi_*$ the maximum value attained by $\xi$ while the roll of the axion is fastest, which occurs when $\chi \simeq \frac{\pi}{2} \,f$ (the small time variation of $H$ can be disregarded for these considerations). The highly amplified gauge fields source perturbations of the axion (via inverse decay) as well as inflaton perturbations and gravitational waves (through gravitational interaction) 
\begin{equation}
    \delta A+\delta A \to \delta \chi \;, \delta \phi ,\, h_\lambda \;\;,
\end{equation}
where $\lambda = \pm$ denotes the two circular GW polarizations. In turn, the axion perturbations can ``convert'' into inflaton perturbations through their linear coupling due to gravity (which arises while the two fields are both rolling). Due to its significant mass during inflation, the energy density in the axion rapidly dilutes away after the field has reached the minimum, so that the primordial scalar density perturbations $\zeta$ in this model can be identified with the inflaton perturbations (more precisely, we work in spatially flat gauge, in which $\zeta = - \frac{H}{\dot{\phi}} \, \delta \phi$). The choice of a decoupled axion and inflaton potential in (\ref{model-U1}) guarantees a minimum amount of production of the primordial scalar perturbations, which was one of the motivations of~\cite{Namba:2015gja} for constructing this model. 
%
\begin{figure}
    \centering
\includegraphics[scale=0.5]{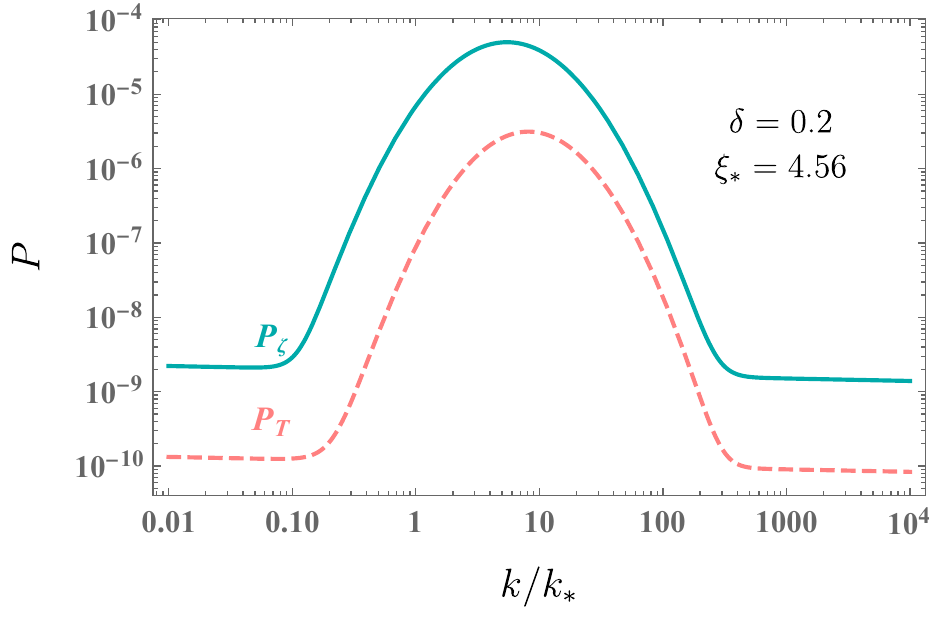}
\caption{
Scalar (solid blue line) and tensor (dashed pink line) power spectra produced in the model (\ref{model-U1}) in which an axion, coupled to a U(1) gauge field, experiences a momentary phase of significant roll during inflation. The results are shown for $\xi_* = 4.56$ (the peak value of the parameter that controls the gauge field production) and $\delta = 0.2$ (the inverse of this parameter is approximately the number of e-folds during which the axion roll is significant). The wavenumber $k_*$ is a model parameter.}  \label{fig:P_together}
\end{figure}

The scalar and tensor perturbations sourced by the gauge field add up incoherently to the standard vacuum modes (those amplified by the inflationary expansion), so that the total scalar and tensor power spectra, namely 
\begin{equation}
    P_\zeta(k)\delta^{(3)}(\vec{k}+\vec{k'}) \equiv \frac{k^3}{2\pi^2}\langle \zeta(\vec{k})\zeta(\vec{k'})\rangle\,,
\label{P-zeta}
\end{equation}
(and analogously for $h_\pm$) are obtained from the sum of the vacuum and sourced contributions 
\begin{equation}
    P_i(k)=P_i^{(0)}(k) +P_i^{(1)}(k) \quad, \, i=\zeta,\,h_+, \, h_-\,.
\end{equation}

We employ the standard parametrization of the scalar vacuum power spectrum:
\begin{equation}
P_\zeta^{(0)}=A_s\left(\frac{k}{k_0}\right)^{n_s-1}\,, 
\end{equation}
and we assume that the axion is still (nearly) at rest when the CMB modes were produced, so that the CMB data are directly mapped only to the vacuum signal. We then take $A_s = e^{3.047}\times10^{-10}$, $n_s = 0.9665$ at the Planck pivot scale $k_0 = 0.05 \, {\rm Mpc}^{-1}$, according to the central values given in Table $2$ of~\cite{Planck:2018vyg}. We also take $\epsilon=0.002$ for the slow roll parameter $\epsilon \equiv \frac{\mpl ^2}{2} \left( \frac{1}{V} \, \frac{d V}{d \phi} \right)^2$ (where $\mpl$ is the reduced Planck mass), which saturates the current bound $r =16\epsilon \lesssim 0.03$ on the tensor-to-scalar ratio~\cite{Tristram:2021tvh,Galloni:2022mok}.
For simplicity, we also take this ratio to be scale independent, 
\begin{equation}
P_T^{(0)}=r \times P_\zeta^{(0)}\,, 
\label{eq-r}
\end{equation}
at all (comoving) wavenumbers $k$. Relaxing this assumption (that is immediately done for any specific choice of the inflationary potential, given the two different dependences of the scalar and tensor tilt on the slow-roll paramters) has no consequence on our results, since in slow-roll inflation the vacuum tensor modes are too small to generate appreciable SDs. 

The rolling of the axion that takes place after the CMB modes have been produced leads to a  bump in the scalar and tensor power spectra at scales smaller than the CMB ones, that can be parametrized as~\cite{Namba:2015gja}
\begin{align}
P_\zeta^{(1)} \left( k \right) &=\left[\epsilon \, P_\zeta^{(0)} \left( k \right) \right]^2 f_{2,\zeta} \;\;\;,\;\;\; 
P_T^{(1)} \left( k \right) =\left[\epsilon \, P_\zeta^{(0)} \left( k \right) \right]^2 \left( f_{2,+}+f_{2,-} \right)\,, 
\end{align}
where the dimensionless functions $f_{2,j}$ are well fitted by~\cite{Namba:2015gja} 
\begin{equation}
    f_{2, j}\simeq f_{2, j}^{c}\left[\xi_{*}, \delta\right] \exp \left[-\frac{1}{2 \sigma_{2, j}^{2}\left[\xi_{*}, \delta\right]} \ln ^{2}\left(\frac{k}{k_{*} x_{2, j}^{c}\left[\xi_{*}, \delta\right]}\right)\right]   \;\;,\;\; j = \zeta ,\, h_+ ,\, h_- \,.
\label{f2j}
\end{equation}
The three functions $f_{2, j}^{c} ,\, \sigma_{2, j}^{2} ,\, x_{2, j}^{c}$ control, respectively, the height, the width, and the location of the bump. Ref.~\cite{Namba:2015gja} provided an analytic fit for the dependence of these functions on $\xi_*$, for the two specific values $\delta = 0.2 ,\, 0.5$. In the present work we only show outcomes for the $\delta = 0.2$ case, as the two choices give qualitatively similar results. The quantity $k_*$ in eq.~(\ref{f2j}) is the comoving wavenumber of the modes that exit the horizon while the axion roll is the fastest. As $x_{2, j}^{c}$ is an order one quantity, this is parametrically the scale at which the sourced modes have maximum amplitude. The scale $k_*$ is a free parameter of the model, which we vary in the following analysis of the SDs produced by the scalar and tensor modes. 

In \cref{fig:P_together} we show the scalar and tensor power spectra generated in this model, for  $\xi_* = 4.56$~\footnote{As the present analysis shows, this is the largest value of $\xi_*$ that is compatible with the current bounds on the distortions for all values of $k_*$, see  \cref{fig:all_dist_2}. As shown in Ref.~\cite{Peloso:2016gqs}, the primordial perturbations produced by this mechanism are under perturbative control for $\xi \la 5$.} (for the $\delta = 0.2$ choice considered here). We note indeed the presence of peaked sourced signals on top of the baseline of the vacuum modes. We also observe that, for our example, the sourced tensor modes at best can be of the order of magnitude of the sourced scalar modes. As a consequence, the CMB distortions produced by both type of modes need to be considered to perform a complete phenomenological study of the model.

\begin{figure}
    \centering

    \includegraphics[scale=0.5]{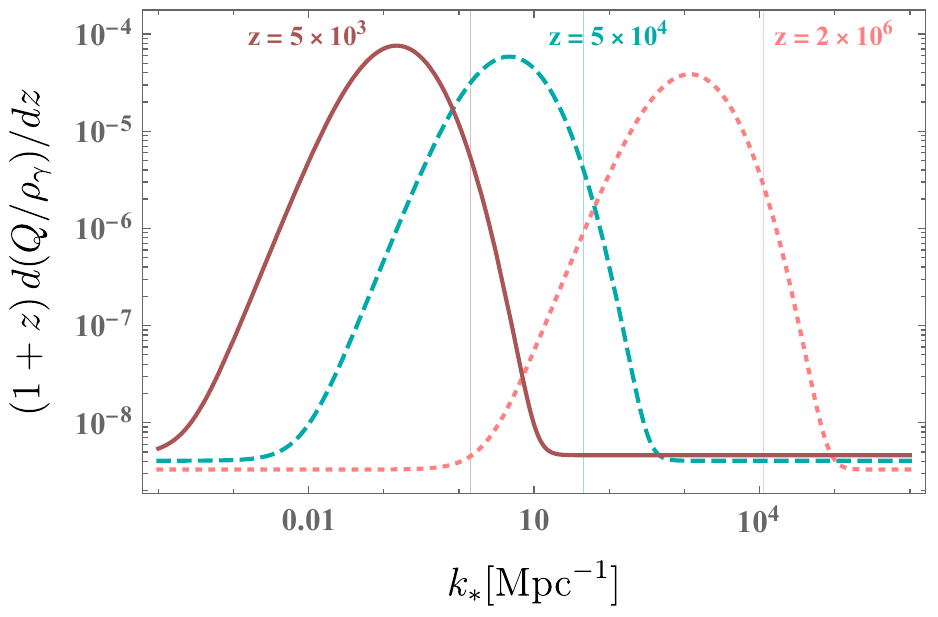}\includegraphics[scale=0.5]{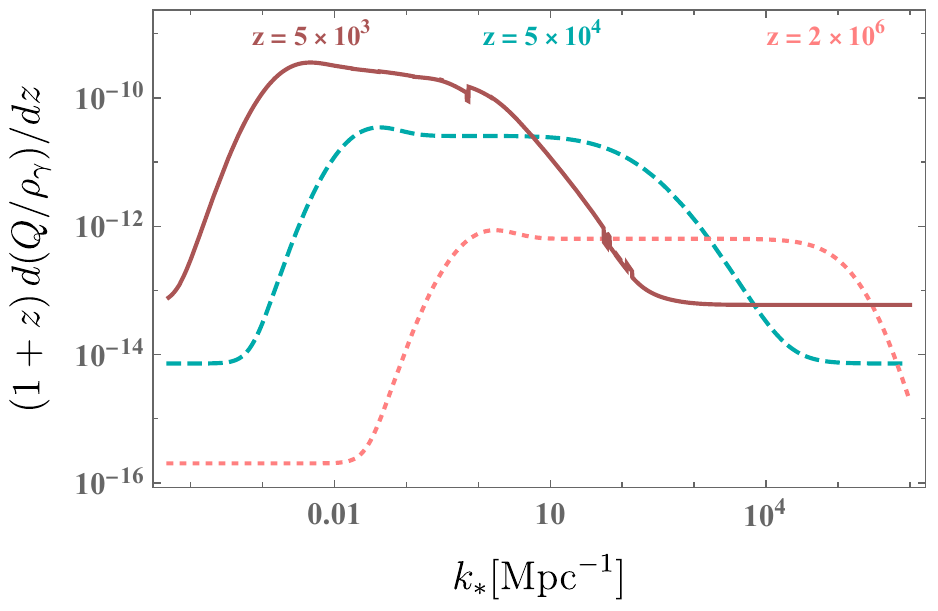}
\caption{
Left (respectively, right) panel: scalar heating rate~(\ref{Q-zeta}) (respectively, tensor heating rate~(\ref{Q-T})) due to the primordial perturbations produced in the model~(\ref{model-U1}) and shown in Figure~\Ref{fig:P_together}. The rates are shown as a function of $k_*$ (which is the comoving momentum, of the modes that leave the horizon when the evolution of the axion is fastest). In each panel, the curves, from left to right, show the heating rates at three increasing redshift: $z=5 \times 10^3$ (solid line), $z=5 \times 10^4$ (dashed line), and $z=2 \times 10^6$ (dotted line). In the left panel the thin vertical lines correspond to the damping wavenumber $k_D(z)$ for each redshift, beyond which the distortions from scalar modes are reduced.}
\label{fig:heatingrates_2}
\end{figure}

We begin this analysis with the scalar and tensor heating rate, shown in \cref{fig:heatingrates_2}. We plot the heating rates (obtained after integration over $k$) as a function of $k_*$ for various relevant redshifts. We observe that, the greater the redshift, the greater is the value $k_*$ at which the heating rate is maximum. This is due to the fact that modes mostly contribute to the heating when they re-enter the horizon, and that 
the comoving horizon is smaller in the past (hence, greater $k = \left( a H \right)^{-1}$). We also observe that, at any given redshift,  the scalar heating rate strongly decreases at scales that are smaller than the diffusion scale $k_D \left( z \right)$ at that redshift. A similar effect is not found for the tensor heating rate, which shows 
a plateau that extends for a  greater range of wavenumbers. This agrees with the large-$k$ behaviour of the window functions shown in \cref{fig:Wcomparison_ap} and with the discussion we had after eq.~(\ref{F_mu_y_T}). 
We also observe that the heating rate caused by tensor modes decreases at high redshifts. On the contrary, the heating rate due to scalars reaches a similar maximum level across the three redshifts examined. Since $\mu-$distortions occur at higher redshifts compared to $y-$distortions, the diminishing effect of tensor heating at high redshifts implies that the ratio between the peak amplitudes of $\mu-$ and $y-$distortions is smaller for tensor contributions compared to scalar contributions, as Figure \ref{fig:all_dist_2} shows. 

\begin{figure}
    \centering
    \includegraphics[scale=0.5]{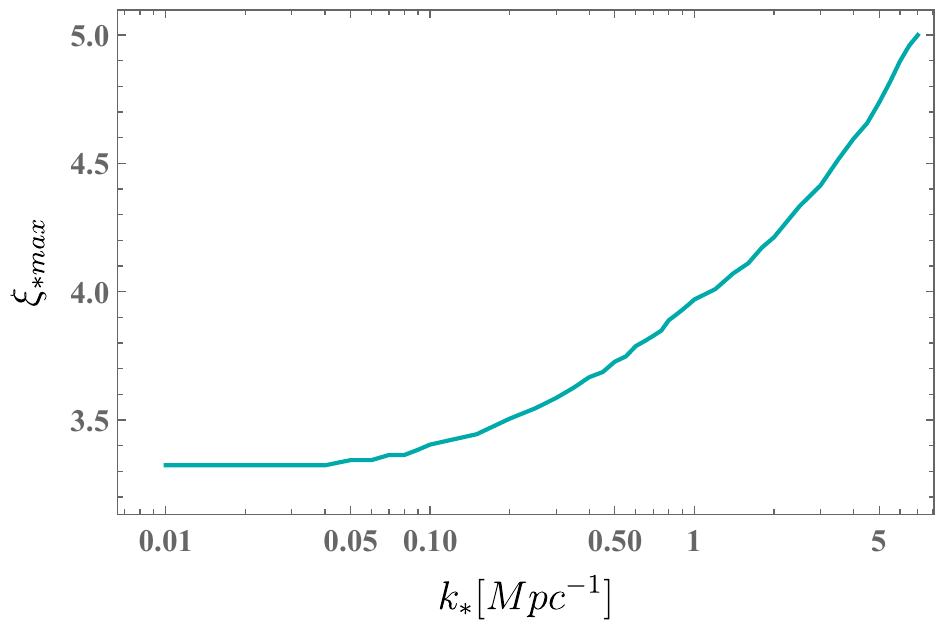}
\caption{
Maximum allowed value of $\xi_*$ as a function of $k_*$. Parameters above the curve result in a sourced power spectrum that exceeds the one inferred from CMB~\cite{Planck:2018jri}.}
\label{fig:Planckxi}
\end{figure}

\begin{figure}
\centering
\includegraphics[scale=0.48]{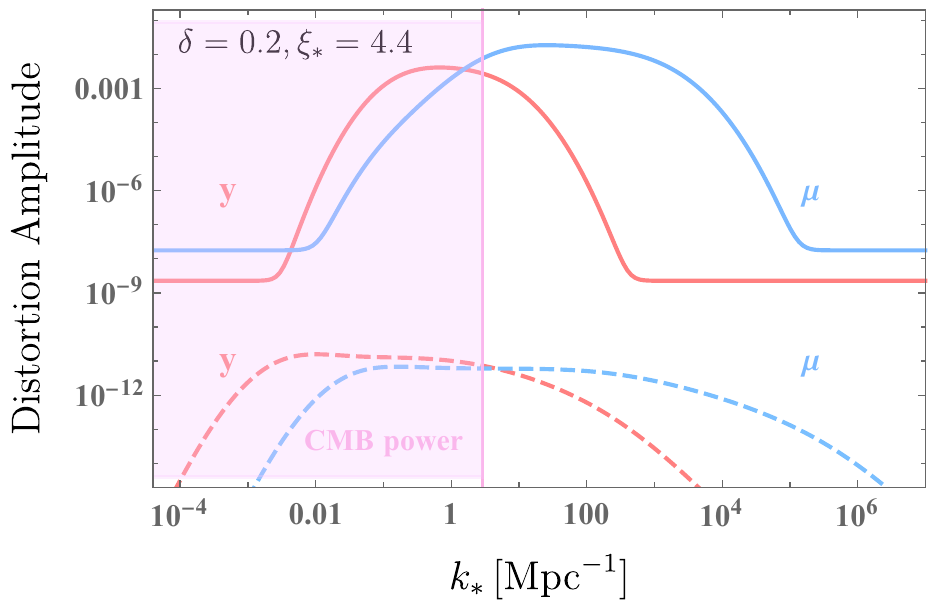}
\includegraphics[scale=0.48]{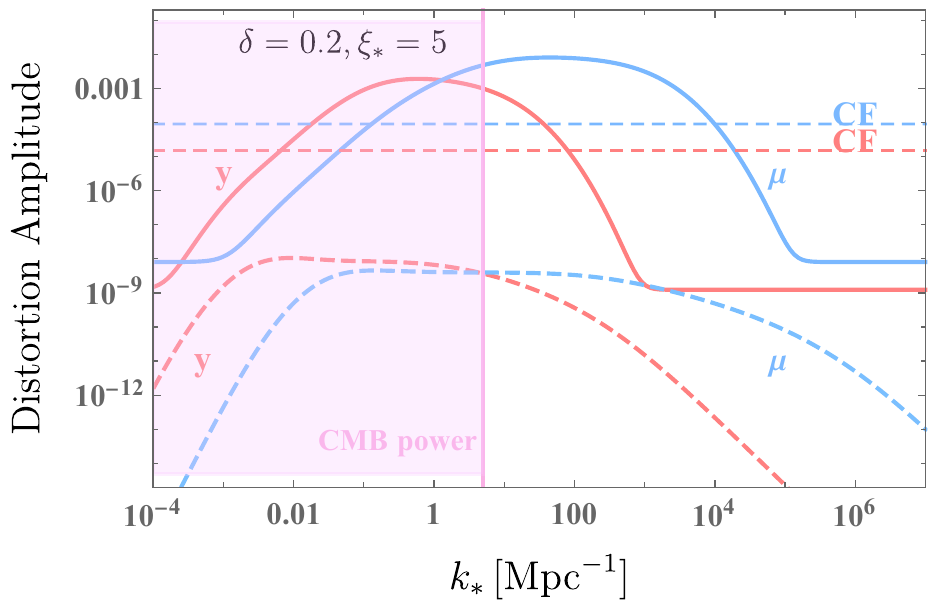}
\caption{
Solid (respectively, dashed) lines: SDs from the scalar (respectively, tensor) perturbations produced by the model 
(\ref{model-U1}) and shown in \cref{fig:P_together}.
The distortions are shown as a function of the parameter $k_*$, which controls the wavenumber at which the  perturbations are peaked. The horizontal lines correspond to the current bounds from COBE/FIRAS~\cite{Mather:1993ij}. The two panels correspond to different values of the parameter $\xi_*$, that controls the amplitude of the sourced perturbations.
The shaded region in the left part of the two panels show the parameters that result in a sourced power spectrum that exceeds that inferred from the CMB (see the previous figure). The flat tails at small and large $k_*$ are due to the vacuum modes and the choice of parameters discussed above~\cref{eq-r}.}
\label{fig:all_dist_2}
\end{figure}

We note that the CMB data already constrain part of the parameter space by placing strong bounds on the strength of the curvature power spectrum up to $k_*\sim 5$ Mpc$^{-1}$. \cref{fig:Planckxi} shows the maximum value of $\xi_*$ as a function of $k_*$ that satisfies the  $2 \sigma$ bound reported in Figure 20 (bottom panel) of Ref.~\cite{Planck:2018jri}. We note that the limit loosens (greater values of $\xi_*$ are allowed) for progressively increasing $k_*$, as this corresponds to progressively smaller scales (eventually becoming smaller than the scales probed by the CMB). This limit excludes the left portion of~\cref{fig:all_dist_2}, which is shaded in the Figure.

Finally, in support of our earlier claim that computing only the tensor contribution would significantly underestimate the CMB distortions from this mechanism, we observe that the scalar heating rate reaches values that are much greater than the tensor ones. This is also in agreement with the window functions shown in \cref{fig:Wcomparison_ap} and with the fact that the tensor modes produced by this mechanism are smaller than the scalar ones.~\footnote{Although we do no perform a full parameter search for the model, we do not expect to obtain tensor modes of many orders of magnitude greater than the scalar ones, so to compensate the hierarchy in the scalar vs. tensor window functions shown in \cref{fig:Wcomparison_ap}, due to the fact that the gauge field is coupled with comparable (gravitational) magnitude to both sectors.}

%
%
%

The SDs for this specific model are shown in \cref{fig:all_dist_2}.~\footnote{We stress that \cref{fig:heatingrates_2} and \cref{fig:all_dist_2} are not spectra, but they show the heating rates and the distortions after integration over $k$. The horizontal axis $k_*$ is the comoving momentum at which the spectra of the primordial perturbations are peaked.} In agreement with the above discussions and results, we observe that indeed the distortions due to tensor modes produced in this model are much smaller than the scalar-induced ones. We also observe the relative size between the maximum value attained by the $\mu-$ vs. $y-$distortions is greater for the scalar contribution than for the tensor one, in agreement with what discussed in the context of the previous figure. Finally, we note that, after the peak, the tensor contributions have a milder decrease with $k_*$ than the scalar ones, and so they reach the baseline minimum value (due to the nearly scale invariant vacuum modes) at greater values of $k_*$. This is also something that we already discussed above. 

In conclusion, the model can result in significant CMB distortions. As already discussed, the amplitude of the signal is exponentially proportional to the parameter $\xi_*$. The left panel of \cref{fig:all_dist_2} corresponds to a value of $\xi_*$ that is compatible with the present limits for all choices of $k_*$. The right panel shows instead the results for the maximum estimated value of $\xi$ that is compatible with perturbativity (in this case the range $6 \times 10^{-3} {\rm Mpc}^{-1} \la k_* \la 10^{4} \,  {\rm Mpc}^{-1}$ is already excluded by COBE/FIRAS). We stress that in all cases, the distortions produced in this model are  strongly dominated by the primordial scalar modes.

\section{Spectator Chromo Natural Inflation}
\label{SU}

Let us now move our attention to the model of ref. \cite{Dimastrogiovanni:2016fuu}, which is essentially an analogous version of the model of the previous section, with the U(1) field replaced by an SU(2) multiplet. We divide our study in three subsections. In Subsection~\ref{subs:SU(2)pert} we present the model and the expressions for the primordial tensor and scalar modes produced by it. In Subsection \ref{subs:SU(2)bckground} we then study the background evolution in this model for some specific examples, constructed to have peaked signals at scales smaller than the CMB ones. In Subsection~\ref{subs:SU(2)dist} we finally present the CMB SDs obtained in these specific examples.

\subsection{Sourced tensors and scalars}
\label{subs:SU(2)pert}

We denote this model as `Spectator Chromo Natural Inflation', since it is also analogous to Chromo Natural Inflation \cite{Adshead:2012kp}, with a spectator axion field, rather than the inflaton, coupled to the SU(2) multiplet. With this choice, the inflationary background dynamics is not controlled by the axion and gauge fields. The model is governed by the action 
\begin{equation}
    \mathcal{L}= -\frac{1}{2}(\partial \phi)^2-V(\phi) - U(\chi)-\frac{1}{2}(\partial \chi)^2-\frac{1}{4} F_{\mu \nu}^a F^{a \mu \nu}+\frac{\lambda \chi}{4 f} F_{\mu \nu}^a \tilde{F}^{a \mu \nu} \,,
\label{L-SU2}
\end{equation}
where $\phi$ is the inflaton and $\chi$ is the spectator axion, coupled to the SU(2) field of field strength 
\begin{equation}
F_{\mu\nu}^a = \partial_\mu A_\nu^a - \partial_\nu A_\mu^a + g f^{abc} A_\mu^b A_\nu^c \;, 
\end{equation}
where $g$ is the gauge coupling constant and $f^{abc}$ are the structure constants of the SU(2) algebra. As in the previous model, the inflaton and the axion-gauge sector are decoupled (apart from gravity), and therefore the inflaton and axion have the separate potentials $V(\phi)$ and $U(\chi)$, respectively. 
The SU(2) triplet has vacuum expectation values (vevs)~\cite{Adshead:2012kp}
\begin{equation}
    A_0^a=0, \quad \quad A_i^a=\delta_i^a a(t) Q(t)\,, 
\label{vevs}
\end{equation}
which allows for a statistically isotropic phenomenology. 

One component of the SU(2) multiplet is strongly amplified due the rolling of the axion, similar to the instability of the U(1) field studied in the previous section. The instability is controlled by the combination
\begin{equation}
m_Q \equiv \frac{gQ(t)}{H_{\rm inf}} \;, 
\label{mQ}
\end{equation}
which plays an analogous role to the parameter $\xi$ introduced in the U(1) case. Indeed, it can be shown that $m_Q \simeq \xi$ in the $m_Q \gg 1$ limit, see eq. (\ref{Q-chidot-sol-slow}) below. 

Differently from the U(1) case, the amplified SU(2) component is linearly coupled to one GW polarization, due to the vevs (\ref{vevs}). Therefore, the latter is significantly sourced already at the linearized level, with the power spectrum~\cite{Dimastrogiovanni:2016fuu}
\begin{equation}
    P_{\rm GW,sourced}(k)=\frac{\epsilon_B H^2}{\pi^2 \mpl ^2}\mathcal{F}^2(m_Q)\,,
    \label{TensorPower}
\end{equation}
where $\epsilon_B \equiv \frac{g^2 Q^4}{\mpl ^2H_{\rm inf}^2} = \frac{m_Q^4H_{\rm inf}^2}{g^2 \mpl ^2}$ is a parameter that roughly indicates the energy fraction of the $SU(2)$ gauge fields. The function $\mathcal{F}(m_Q)$ was evaluated in Ref.~\cite{Dimastrogiovanni:2016fuu}, and reported in their eqs. (3.6)-(3.9). This is the expression that we use in our computations. Ref.~\cite{Thorne:2017jft} obtained a much simpler approximate fitting function: 
\begin{equation}
\mathcal{F} (m_Q) \simeq {\rm exp} 
\left[ 2.4308 \, m_Q - 0.0218 \, m_Q^2 - 0.0064 \, m_Q^3 - 0.86 \right] 
\;\;,\;\; 3 \leq m_Q \leq 7 \;. 
\label{fit-F-mQ}
\end{equation}

As in the U(1) model, also the scalar perturbations need to be computed to provide the phenomenology of the model.  The power spectrum of the linear modes is given by (see Appendix~F of~\cite{Papageorgiou:2019ecb} for a derivation)
\begin{equation}
    P_\zeta^{(0)} \simeq \frac{H^2}{8 \pi^2 M_p^2}\frac{\epsilon_\phi}{(\epsilon_\phi+\epsilon_B)^2} \simeq 
    P_{\zeta,{\rm CMB}} \cdot \frac{\epsilon_\phi^2}{(\epsilon_\phi+\epsilon_B)^2}\,,
\label{Pze0}
\end{equation}
where $\epsilon_\phi \equiv \frac{\dot{\phi}^2}{2 H^2 \mpl ^2}$ is the standard slow-roll parameter associated with the motion of the inflaton. As in the previous section, we consider a model characterized by negligible axion roll when CMB perturbations are produced, and therefore standard vacuum signals at CMB scales, which justifies the last expression in eq.~(\ref{Pze0}), where $P_{\zeta,{\rm CMB}}$ is the power observed at CMB scales, with the amplitude $A_s$ reported in the previous section. We employ the results of Ref.~\cite{Papageorgiou:2019ecb}, which, in addition to the scalar perturbations from linear theory, also computed the scalar perturbations sourced at nonlinear order by two amplified gauge modes. The power spectrum of these sourced scalar modes is approximately given by
\begin{equation}
P_{\zeta,{\rm sourced}} \simeq 
1.1 P_{\zeta,{\rm CMB}}^2 \frac{\epsilon_\phi^4}{(\epsilon_B+\epsilon_\phi)^2}  N_k^2  m_Q^{11} \text{e}^{7m_Q}  \;\;,\;\; 2.5 \lesssim m_Q \lesssim 4 \;, 
\label{Pze1}
\end{equation}
where $N_k$ is the number of inflationary e-folds during which the axion rolls. 

While Ref.~\cite{Dimastrogiovanni:2016fuu} studied this model in a regime in which the axion rolls continuously all throughout inflation, Ref.~\cite{Thorne:2017jft} was interested in the case in which the axion moves significantly for only a few e-folds during inflation, analogous to the model considered in the previous section.~\footnote{ 
The implementation of~\cite{Thorne:2017jft} was then employed by~\cite{Kite:2020uix} as a model for SDs from tensors, which is the main motivation for this section. Strictly speaking, the relations (\ref{TensorPower}) and~(\ref{fit-F-mQ}) used in these two works for the tensor signal, as well as the relation (\ref{Pze1}) for the source scalars were obtained for constant $m_Q$, and then evalauted for a time varying $m_Q$ in Refs.~\cite{Thorne:2017jft,Kite:2020uix} (specifically, for any wavenumber $k$, the relations~(\ref{TensorPower}) and~(\ref{fit-F-mQ}) are evalauted at the time at which that mode left the horizon). As our goal is to compare the results of~\cite{Kite:2020uix} for the case in which also the scalar modes are considered, we employ in this work the same methodology used in Refs.~\cite{Thorne:2017jft,Kite:2020uix}. However, we caution that the relations.~(\ref{TensorPower}), (\ref{fit-F-mQ}) and (\ref{Pze1}) can only be considered as approximations in the case in which $m_Q$ varies too quickly.} Ref.~\cite{Thorne:2017jft} did not study a concrete dynamical evolution for the axion in a specific model. They 
considered the typical axion potential (\ref{Vaxion-cos}) and Taylor expanded the evolution of the axion about the time $t_*$ when it is at the steepest part of the potential, and therefore $\dot{\chi} \left( t = t_* \right) \equiv \dot{\chi}_*$ is maximum, 
\begin{equation}
\chi \left( t \right) = \frac{\pi}{2} \, f + \dot{\chi}_* \left( t - t_* \right) + {\rm O } \left( \left( t - t_* \right)^2 \right) \;. 
\label{chi(t)Taylor}
\end{equation}
Then, inserting this expansion in eqs.~(\ref{TensorPower}) and~(\ref{fit-F-mQ}), they obtained an analytically approximate expansion for the top of the bump of the tensor modes produced while the axion is rolling. 

As we study in the next subsection, realizing such a bump in the SU(2) case is more difficult than in the U(1) model considered in the previous section. This cannot be seen from the analytical approximation~(\ref{chi(t)Taylor}) that only describes the evolution around the moment of fastest roll, but it emerges from the study of the full dynamical background equations in concrete models. We discuss this difficulty in the next subsection, where we outline a possible construction that can indeed produce a fast roll of the axion restricted to the scales that are relevant for the CMB distortions.

\subsection{Background evolution for localized peaks beyond the CMB scales}
\label{subs:SU(2)bckground}

Let us study the background evolution for the model~(\ref{L-SU2}). Concerning the dominant inflaton sector, we consider for definiteness the so called $\alpha$-attractor potential 
\begin{equation}
    V=V_0\tanh\left(\frac{\phi}{\sqrt{6\alpha}\mpl}\right)\,,
\end{equation}
which has the advantage of providing a nearly constant $H$ during inflation, and of admitting an analytic solution for the evolution of the inflaton and of the expansion rate (in the limit of negligible energy in the axion/gauge sector that we are considering), see e.g. Ref.~\cite{Kallosh:2013yoa}. For definiteness, we choose $V_0= 1\times 10^{-9} \mpl ^4$, $\alpha=1$, and $\phi_{\rm in} = 6.24 \mpl$ (leading to $60$ e-folds of inflation). Other inflaton potentials might have been considered without affecting our results.

The background evolution of the axion/gauge sector is instead controlled by 
\begin{eqnarray}
&& \ddot{\chi} + 3 H \dot{\chi} + \frac{d U}{d \chi} = - \frac{3 g \lambda}{f} Q^2 \left( \dot{Q} + H \, Q \right) \;, 
\nonumber\\
&& \ddot{Q} + 3 H \dot{Q} + \left( \dot{H} + 2 H^2 \right) Q + 2 g^2 Q^3 = \frac{g \lambda}{f} \, \dot{\chi} \, Q^2 \;.  
\label{eqbck-SU2}
\end{eqnarray}
In the limit of constant or adiabatically evolving $Q$, and provided the parameters satisfy $3 f^2 H^2 \ll g^2 \lambda^2 Q^4$ and $\lambda^2 Q^2 \gg 2 f^2$, these equations are approximately solved by \cite{Adshead:2012kp,Dimastrogiovanni:2016fuu}
\begin{equation}
Q \simeq \left( - \frac{f }{3 g \lambda H} \, \frac{d U}{d \chi} \right)^{1/3} \;\;,\;\; \frac{\lambda \dot{\chi}}{2 f H} \simeq m_Q + \frac{1}{m_Q} \;. 
\label{Q-chidot-sol-slow}
\end{equation} 

We are not interested in a solution with constant $Q$, as we want to obtain a peaked signal. As shown in \cref{fig:Wcomparison_ap}, the tensor window function can dominate over the scalar one at relatively smaller scales. Consequently, an ideal scenario for obtaining visible distortion from tensor modes involves tailoring the model to produce a peak in $m_Q$ at these relatively small scales, effectively minimizing the scalar contribution. 

\begin{figure}
    \centering
    \includegraphics[scale=0.45]{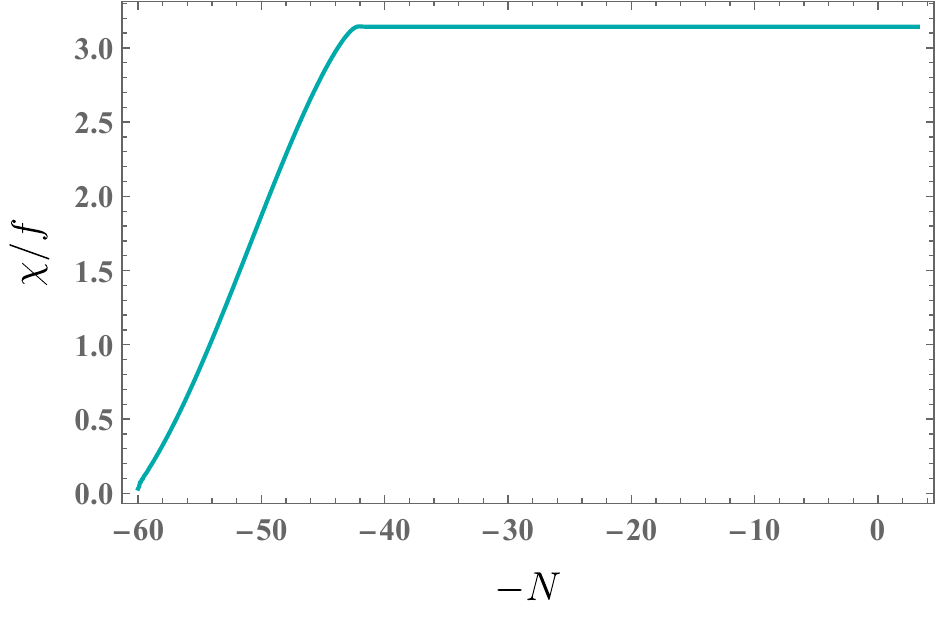}
     \includegraphics[scale=0.45]{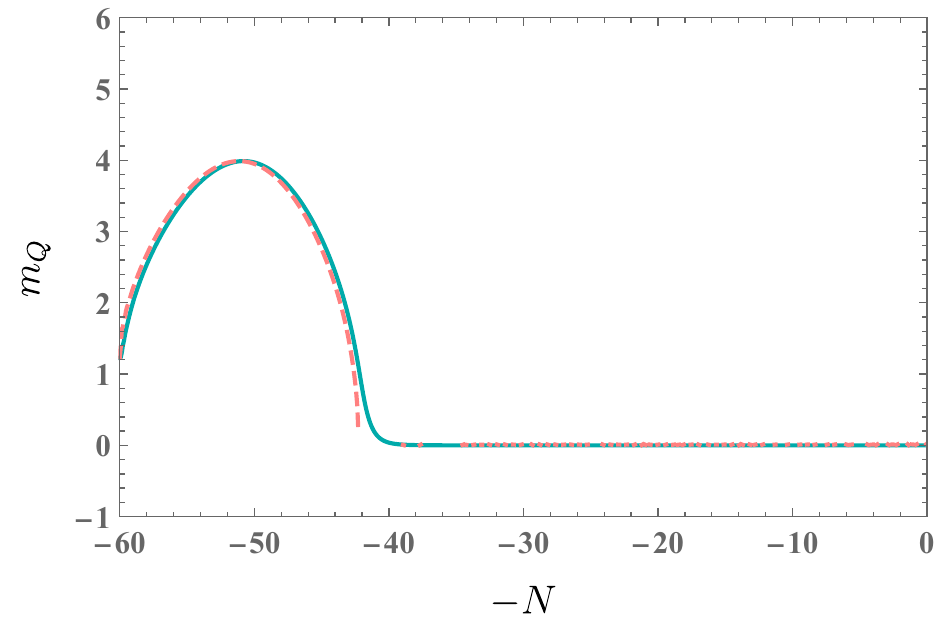}
    \caption{Evolution of the axion $\chi$ (left panel) and of the combination $m_Q$ (right panel) in the SU(2) case for the values of parameters indicated in the main text. Continuous blue lines show the exact numerical solution. The dashed red line in the right panel is based on the approximation of \cref{Q-chidot-sol-slow}, and it presents gaps after the peak since this approximation results in an imaginary $Q$ whenever $\chi > \pi f$ (where the approximation is invalid).}
    \label{fig:N1}
\end{figure}

The tensor (as well as the scalar) signal is generated by the gauge fields, whose amplification is directly linked to the velocity of the axion. For the cosine potential (\ref{Vaxion-cos}), using the relations (\ref{Q-chidot-sol-slow}) in the $m_Q \gg 1$ limit, we can estimate the value of $m_Q$ at the peak 
\begin{equation}
m_{Q,{\rm peak}} \sim \left( \frac{g^2 \, \Lambda^4}{3 \lambda \, H^4} \right)^{1/3} \;, 
\label{mQpeak-est}
\end{equation}
as well as the width of the peak (namely, the number of e-folds for which the axion rolls) 
\begin{equation}
\Delta N_{\rm peak} \sim H \, \Delta t \sim H \, \frac{\Delta \chi}{\dot{\chi}} \sim \frac{\pi}{2} \frac{\lambda}{m_Q} 
\sim {\rm O } \left( 2 \right) \times \left( \frac{\lambda \, H}{\sqrt{g} \, \Lambda} \right)^{4/3} \;. 
\label{DN-est}
\end{equation} 
The order two factor in the last expression comes from the numerical factors and from the fact that we have estimated $m_Q$ at its maximum (at $\chi = \frac{\pi}{2} f$), while in fact $m_Q$ is smaller at the beginning and at the end of the peak. For definiteness, we choose $g= 5.5 \times 10^{-3}$, $\lambda=40$, $f= 9\times 10^{-3}\mpl$, $\Lambda =2.3\times 10^{-3} \mpl$, and $\chi_{\rm in} = 10^{-2} \pi \, f$. This leads to $\Delta N_{\rm peak} \sim 10$ in the estimate (\ref{DN-est}), and it ensures that the conditions leading to (\ref{Q-chidot-sol-slow}) are satisfied. 

In \cref{fig:N1} we show the evolution of the axion and of the parameter $m_Q$ (\ref{mQ}) for this choice of parameters. As mentioned, we start from $\chi_{\rm in}=10^{-2}\pi \, f$, and the initial conditions for $\dot{\chi}$ and $Q$ are then set according to (\ref{Q-chidot-sol-slow}), while $\dot{Q}_{\rm in} = 0$). This results in a bump of $\dot{\chi}$, and hence of $m_Q$ (and a consequent peak in the primordial scalar and tensor perturbations) right at the start of the evolution, as obtained from an exact numerical integration of eqs.~(\ref{eqbck-SU2}), shown by a continuous blue line in the figure. At the beginning of the evolution and all throughout the bump the approximation (\ref{Q-chidot-sol-slow}), indicated by a dashed  red line, is well satisfied. 

We notice that the bump of particle production occurs when $\chi = \frac{\pi}{2} \, f$, and the axion roll is fastest. As mentioned, we wish to delay the moment at which this takes place, so to have the bump at scales smaller than the CMB one. Achieving this in the U(1) model of Section \ref{U1} is rather simple, as it only requires taking the initial value of the axion sufficiently close to the maximum of the potential, where the potential is flat and therefore the classical value of $\dot{\chi}$ can be made arbitrarily small. This is no longer the case now, due to the nonvanishing right hand side in the first of eqs.~(\ref{eqbck-SU2}). We performed an evolution of the system for $\chi_{\rm in}=10^{-4}\pi \, f$ (without changing the other parameters) obtaining a qualitatively similar evolution to the one shown in \cref{fig:N1}, with no delay of the onset of the peak. Further decreasing the initial value of $\chi_{\rm in}$ breaks the conditions reported before eq. (\ref{Q-chidot-sol-slow}). 

\begin{figure}
    \centering
    \includegraphics[scale=0.4]{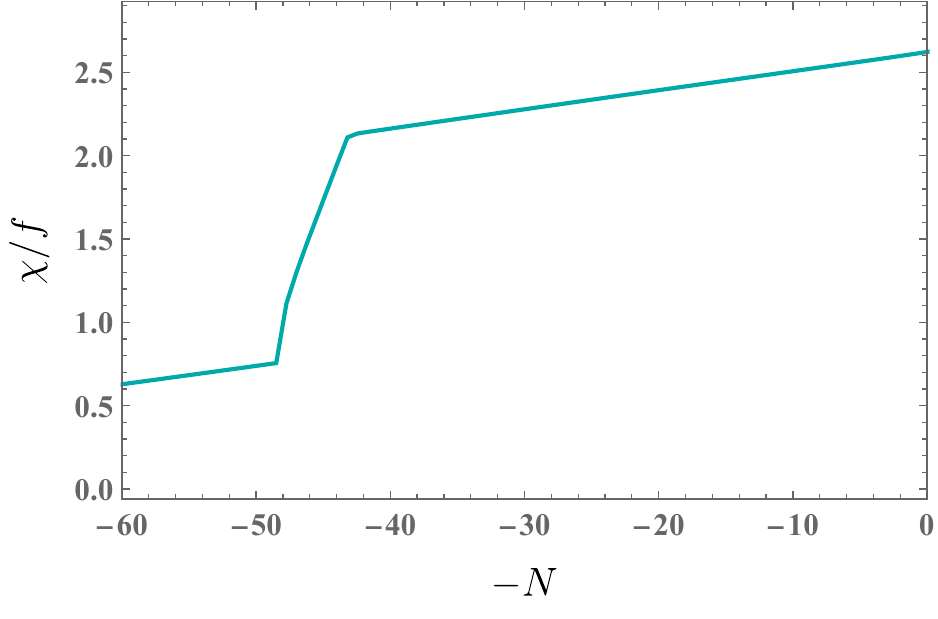}
     \includegraphics[scale=0.4]{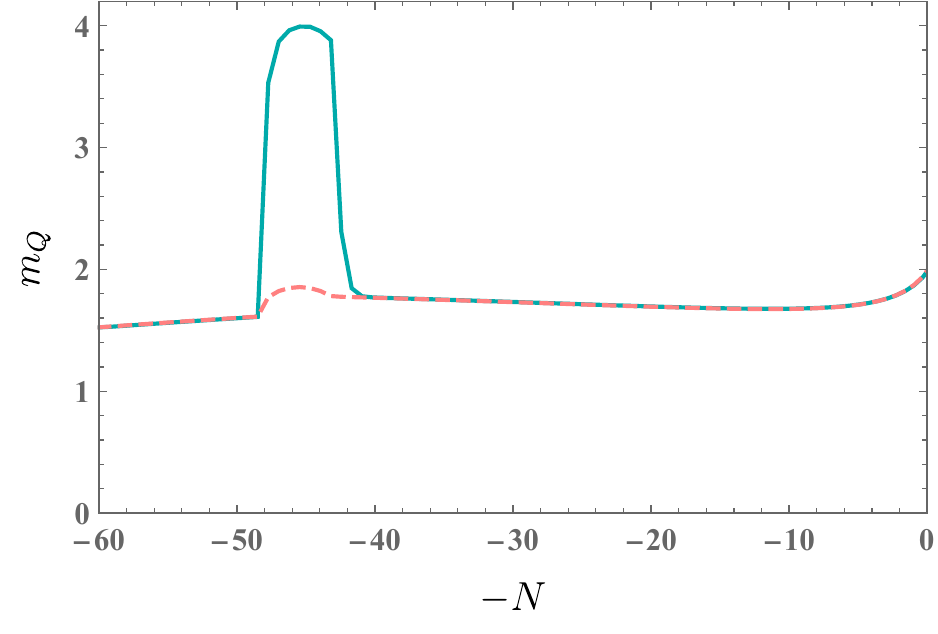}
    \caption{Evolution of the axion $\chi$ (left panel) and of the combination $m_Q$ (right panel). As compared to the previous figure, $\lambda$ is decreased by a factor of $10$ in the interval $48 \gtrsim N \gtrsim 43 $, resulting in peak in $m_Q$ during this stage. The red-dashed curve in the right panel is based on the approximations (\ref{Q-chidot-sol-slow}), which are invalid at the peak.}
    \label{fig:N2}
\end{figure}

Eqs. (\ref{mQpeak-est}) and (\ref{DN-est}) suggest an alternative, and more successful, route to attempt to delay the onset of the peak. We note that the height and the duration of the peak are, respectively, decreasing and increasing as $\lambda$ increases. If therefore $\lambda$ is decreased for an intermediate stage during inflation, a higher and narrower peak in $m_Q$ can be obtained during this stage, on the top of a slowly evolving baseline (that in turn corresponds to the much smaller and wider peak that occurs while $\lambda$ is large). This is visible in the evolution shown in \cref{fig:N2}, where $\lambda$ is initially (at $\chi_{\rm in} = \frac{\pi}{5} f$) taken to be $\lambda=400$, and it is suddenly decreased by a factor of $5$ between $N \simeq 48$ and $N \simeq 43$ e-folds before the end of inflation (more precisely, the evolution in the figure corresponds to $N=48.3$ and $N=43.1$). We are implicitly assuming that the value of $\lambda$ is controlled by some other field, that experiences a sudden transition at the beginning and the end of this stage (presenting a complete model is beyond the scope of this work). 

This successfully delays the onset of the main peak. We note that, for the sole purpose of delaying the peak it is not necessary that $\lambda$ grows back to relatively large values after the peak. However, if this does not occur, $m_Q$ rapidly decreases after the peak, becoming smaller than $\sqrt{2}$. When this happens, the system of scalar perturbations in this model has a strong instability in the sub-horizon regime~\cite{Dimastrogiovanni:2012ew}. The evolution shown in \cref{fig:N2} is exempt from this problem. 

We note that the dashed red line in the right panel of \cref{fig:N2} departs from the solid blue curve during the peak, indicating the approximation (\ref{Q-chidot-sol-slow}) is invalid during this stage. These relations are therefore not used in the computation of the scalar and tensor modes, to which we turn next. 

\subsection{Spectral Distortions for a localized peak in the SU(2) model}
\label{subs:SU(2)dist}

\begin{figure}
    \centering
    \includegraphics[scale=0.4]{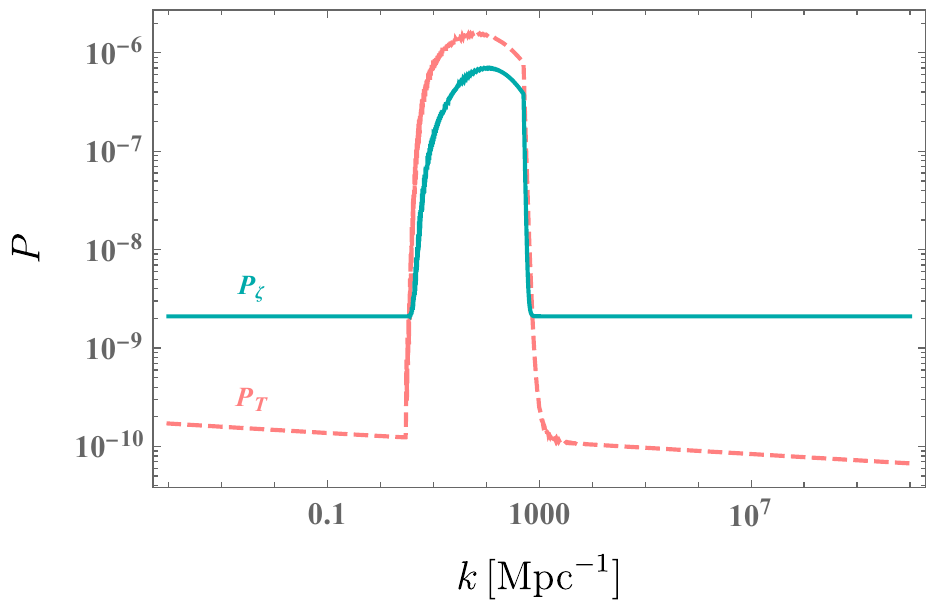}
     \includegraphics[scale=0.4]{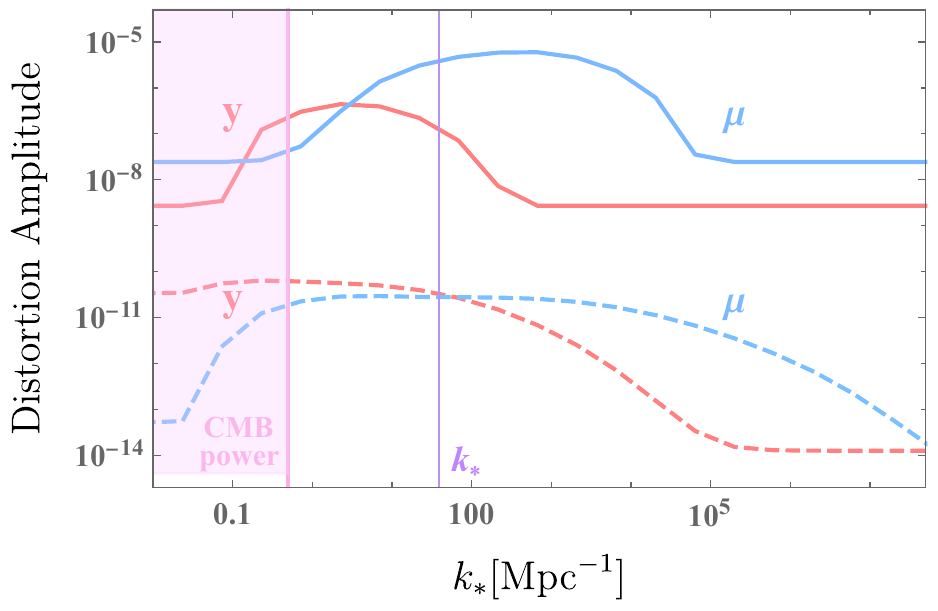}
    \caption{Left panel: scalar (solid line) and tensor (dashed line) power spectrum produced in the model (\ref{L-SU2}) for the background evolution shown in the previous figure. Right panel:
    SDs from the scalar (solid lines) and tensor (dashed lines) modes. The distortions are shown as a function of the parameter $k_*$, which controls the wavenumber at which the  perturbations are peaked. This is varied by shifting the position of the peak shown in the left panel, as explained in the text. The vertical line in the right panel corresponds to the value of $k_*$ obtained for the evolution leading to the spectra shown on the left panel. The shaded region on the left portion of the panel is excluded since it leads to an induced scalar power in excess to what inferred from the CMB.}
    
    \label{fig:P_SU2}
\end{figure}
Let us now compute the primordial scalar and tensor modes produced in the model. We consider the background evolution shown in \cref{fig:N2}. We then evaluate the spectrum of the scalar modes from the sum of eqs.~(\ref{Pze0}) and~(\ref{Pze1}) and the spectrum of the tensor modes from the sum of the vacuum term $P_T^{(0)} = \frac{2}{\pi^2} \, \frac{H^2}{\mpl ^2}$ and of eq. (\ref{TensorPower}). The resulting spectra are shown in the left panel of \cref{fig:P_SU2}. We observe that in this case one can obtain a tensor power spectrum of comparable or greater amplitude than the scalar power spectrum. Inserting the scalar spectrum in eq.~(\ref{mu-y-zeta}) we then obtain the peak values of the CMB distortions $y_\zeta\simeq 3.1\times 10^{-7}$ and $\mu_\zeta\simeq 2.7 \times 10^{-6}$. Inserting the tensor spectrum in eq.~(\ref{mu-y-T}) we instead obtain $y_T\simeq 3.8 \times 10^{-11}$ and $\mu_T\simeq 3.3\times 10^{-11}$. We note that the scalar contribution strongly dominates the distortions, due to their much greater window function, despite the parameters of the model resulting in scalar and tensor modes of comparable power at the peak. 

To shift the position of the peak, we simply change the value of the number of e-folds $N$ at which the decrease of $\lambda$ takes place (we keep the duration of the stage of low $\lambda$ fixed), while all the other parameters are unchanged. We then compute the primordial scalar and tensor modes produced also in these cases, and the corresponding CMB distortions. This allows us to plot in the right panel of \cref{fig:P_SU2} the distortions as a function of the scale of the peak $k_*$. 

We note that the scalar contribution dominates the distortions for all values of $k_*$. The distortions obtained in this model are slightly smaller than those shown in \cref{fig:all_dist_2} in the U(1) case. To increase the distortions generated by this model requires increasing $m_Q$ to values for which the fitting formula~(\ref{Pze1}) is no longer valid. Given the functional dependence of the tensor and scalar signal on $m_Q$, and the fact that the nonlinear scalar production involves more unstable modes than the linear tensor one we do not expect that increasing $m_Q$ would lead to a drastic increase of the ratio between the tensor and scalar sourced modes.

Similar to the $U(1)$ case, the parameter space at relatively small $k_*$ visible in \cref{fig:P_SU2} is excluded as it produces a scalar power spectrum in excess to that inferred by the CMB data. The limit shown in the figure has been computed identically to what done in the previous section for the Abelian case.

%
    \label{fig:planckSU(2)}

\section{Conclusions}\label{sec:conclusions}

CMB SDs are a probe to a plethora of physical processes that might have occurred in the early universe at redshifts below $\simeq 10^6$. One unavoidable source of spectral distortions within the standard cosmological mode is the entropy release associated at the horizon re-entry of primordial perturbations. While the vast majority of studies has focused on scalar perturbations, Refs.~\cite{Chluba:2014qia,Kite:2020uix} computed the distortions generated by primordial tensor modes. This opens an exciting new window in one of the most active current research fields, namely the detection of gravitational waves. 

Primordial tensor modes can be produced by a variety of mechanisms in the early universe (see for instance Refs.~\cite{Bartolo:2016ami,LISACosmologyWorkingGroup:2022jok} for an extensive study of possible cosmological sources at LISA frequencies, many of which can also produce a significant tensor signal at the smaller frequencies that map to CMB SDs). On the other hand, these cosmological mechanisms unavoidably also generate scalar perturbations. Therefore, a complete study of SDs from tensor modes in a specific model needs to be accompanied by the associated study of scalar-induced distortions. In fact, the comparison of the tensor and scalar window functions that we present in \cref{fig:Wcomparison_ap} indicates that the tensor-induced distortions can be overwhelmed by the scalar-induced ones, unless the model can produce a primordial tensor signal that is much larger than the scalar one.~\footnote{Specifically, the ratio between the peak values of the tensor and the scalar window functions shown in the figure amounts to $\frac{W_\mu^T}{W_\mu^\zeta}={\rm O}(10^{-5})$ for the $\mu-$distortions and to $\frac{W_y^T}{W_y^\zeta}={\rm O}(10^{-4})$ for the $y-$distortions. }

Ref.~\cite{Kite:2020uix} studied, among other examples, one such model~\cite{Adshead:2012kp,Dimastrogiovanni:2016fuu,Thorne:2017jft} that is often employed for the generation of a large tensor signal. 
This model is characterized by a rolling axion coupled to an SU(2) gauge field. The coupling significantly generates one component of the SU(2) multiplet, which is linearly coupled to the primordial tensor modes. This, therefore, appears as a natural candidate for generating large and visible tensor signals. However, scalar perturbations are also significantly produced in this model from the nonlinear interactions of the unstable gauge component ~\cite{Papageorgiou:2018rfx,Papageorgiou:2019ecb}. Our findings for the tensor-induced SDs agree with those of Ref.~\cite{Kite:2020uix} that this specific implementation cannot produce significant tensor-induced SDs. Specifically, they obtain a $\mu-$distorion as large as $2 \times 10^{-11}$, which is well reproduced by our~\cref{fig:P_SU2}. Our work expands the findings of Ref.~\cite{Kite:2020uix} by exploring the role of scalar modes, which suggest a more optimistic scenario for future observations. 

In this work we studied the CMB distortions produced by this model, as well as from the earlier U(1) version of~\cite{Namba:2015gja}, including both the contributions from scalar and tensor perturbations. We found that the scalar-induced distortions are significantly greater than the tensor ones, and at a level that can be probed by the proposed probes of SDs (in fact, we also showed some values of parameters in the U(1) case that are already excluded by the COBE/FIRAS data). As already remarked, the response functions to primordial scalar and tensor modes are highly hierarchical. In principle, one might envisage two ways to overcome this hierarchy, so that the tensor-induced SDs are not overpowered by the scalar-induced ones: (1) a mechanism that produces a primordial tensor signal of more than $4$ orders of magnitude greater than the scalar one, and / or (ii) a mechnism that produces a tensor signal at scales relevant for SDs, while a scalar signal on much smaller scales, where the window function is suppressed. Neither case is realized in the models we studied. This is ultimately a consequence of the universality of the gravitational interactions that we have considered: both scalar and tensor modes are only gravitationally coupled to the sourcing gauge fields, and it is not surprising that they do not present strong hierarchies, neither in their amplitude nor in the scales at which they are peaked. It would be interesting to investigate whether this continues to be true beyond standard gravity, as for instance in generalized Galileons inflationary models~\cite{Kobayashi:2011nu}.

In the studied models, the axion rolls for a finite duration (of about $5-10$ e-folds) during inflation, producing a localized bump in the primordial signals. In performing this analysis, we noticed that realizing this localized signal at scales which are relevant for the CMB distortions is more challenging in the SU(2) case than in the U(1) model. We proposed a way out of this problem, involving a time dependence of the axion-gauge coupling. We hope that this idea can be employed also in other applications when this model is set to produce a localized feature in the tensor spectum, as for instance in the CMB study of~\cite{Thorne:2017jft,LiteBIRD:2023zmo}.

\section*{Acknowledgements} 
M. Putti would like to thank J. Leedom and A. Westphal. 
M. Putti is supported by the Deutsche Forschungsgemeinschaft under Germany's Excellence Strategy - EXC 2121 ``Quantum Universe'' - 390833306. S.B. acknowledges support by “Progetto di Eccellenza” of the Department of Physics and Astronomy of the University of Padua. N. Bartolo acknowledges support from Istituto Nazionale di Fisica Nucleare (INFN) through the INFN specific Initiative InDark. N. Bartolo  acknowledges support from the COSMOS network (www.cosmosnet.it) through the ASI (Italian Space Agency) Grants 2016- 24-H.0, 2016-24-H.1-2018 and 2020-9-HH.0. N. Bartolo  acknowledges support from the MUR PRIN2022 Project “BROWSEPOL: Beyond standaRd mOdel With coSmic microwavE background POLarization”-2022EJNZ53  financed by the European Union - Next Generation EU. S. Bhattacharya and M. Peloso acknowledge support from Istituto Nazionale di Fisica Nucleare (INFN) through the Theoretical Astroparticle Physics (TAsP) project. M. Peloso acknowledges support by the MIUR Progetti di Ricerca di Rilevante Interesse Nazionale (PRIN) Bando 2022 - grant 20228RMX4A.

\bibliographystyle{JHEP}
\bibliography{biblio}

\end{document}